\documentclass[aps,prd,preprint,floatfix,preprintnumbers,nofootinbib,superscriptaddress,
showkeys]{revtex4}

\usepackage{hyperref}

\usepackage{graphicx}
\usepackage{bm}
\usepackage{amssymb}
\usepackage{amsmath}
\usepackage{dcolumn}
\usepackage{color}
\usepackage[latin1]{inputenc}
\usepackage[english]{babel}

\begin{document}


\title{Some specific solutions to the translation-invariant $N$-body harmonic oscillator Hamiltonian}

\author{Cintia T. \surname{Willemyns}}
\email[E-mail: ]{cintia.willemyns@umons.ac.be}
\thanks{ORCiD: 0000-0001-8114-0061}

\author{Claude \surname{Semay}}
\email[E-mail: ]{claude.semay@umons.ac.be}
\thanks{ORCiD: 0000-0001-6841-9850}

\affiliation{Service de Physique Nucl\'{e}aire et Subnucl\'{e}aire,
Universit\'{e} de Mons,
UMONS Research Institute for Complex Systems,
Place du Parc 20, 7000 Mons, Belgium}
\date{\today}

\begin{abstract}
\textbf{Abstract} 
The resolution of the Schrödinger equation for the translation-invariant $N$-body harmonic oscillator Hamiltonian in $D$ dimensions with one-body and two-body interactions is performed by diagonalizing a matrix $\mathbb{J}$ of order $N-1$. It has been previously established that the diagonalization can be analytically performed in specific situations, such as for $N \le 5$ or for $N$ identical particles. We show that the matrix $\mathbb{J}$ is diagonal, and thus the problem can be analytically solved, for any number of arbitrary masses provided some specific relations exist between the coupling constants and the masses. 
We present analytical expressions for the energies under those constraints.
\keywords{$N$-body harmonic oscillator Hamiltonian}
\end{abstract}

\maketitle

The general translation-invariant $N$-body harmonic oscillator Hamiltonian in $D$ dimensions with one-body and two-body forces is given by
\begin{equation}
\label{Hho}
H_\textrm{ho} = \sum_{i=1}^N \frac{\bm p_i^2}{2 m_i} -\frac{\bm P^2}{2 M} + \sum_{i=1}^N k_i\, (\bm r_i - \bm R)^2 + \sum_{i<j=2}^N g_{ij}\, (\bm r_i - \bm r_j)^2.
\end{equation}
The momentum $\bm p_i$ of the $i$th particle of mass $m_i$ is the conjugate variable of its position $\bm r_i$. The center of mass coordinate is noted $\bm R = \sum_{i=1}^N m_i \,\bm r_i/M$ where $M=\sum_{i=1}^N m_i$, and the total momentum is noted $\bm P = \sum_{i=1}^N \bm p_i$. The Schrödinger equation for this Hamiltonian can be solved because the Hamiltonian can be rewritten as a sum of $N-1$ decoupled harmonic oscillators
\begin{equation}
\label{Eohint}
H_\textrm{ho}=\sum^{N-1}_{i=1} \left[ \frac{\bm \sigma_i^2}{2m} + \frac{1}{2} m\, \omega_i^2 \bm z_i^2 \right],
\end{equation}
where $m$ is an arbitrary mass scale ($m$ can be one of the masses of the system or $M$, for instance), $m_i=m\,\alpha_i$ for $i=1,2,\ldots,N$. The $\bm z_i$ and $\bm \sigma_i$ are new conjugate variables resulting from a change of variables defined in \cite{silv10} where the center of mass reference frame has been adopted. 


The energy of the system is then given by ($\hbar = 1$)
\begin{equation}
\label{Eohgen}
E_\textrm{ho}=\sum^{N-1}_{i=1} \omega_i\,Q_i,
\end{equation}
where $Q_i = n_i+1/2$ for $D=1$ and $Q_i = 2n_i+l_i+D/2$ for $D\ge 2$ \cite{yane94}. The $n_i, l_i$ are the quantum numbers associated with the coordinates of the harmonic oscillators in Eq.~(\ref{Eohint}). The coefficients $\omega_i$ are given by $d_i=m\,\omega_i^2/2$ where $d_i$ are the eigenvalues of a symmetrical matrix of order $N-1$, let us say $\mathbb{J}$. This matrix can be written as $\mathbb{J}=\mathbb{F}+\mathbb{G}$ where each term corresponds to the contributions from the one-body and two-body interactions respectively. The matrix elements of $\mathbb{F}$ are given in \cite{silv10} and those of $\mathbb{G}$ in \cite{silv10,ma00}, and they can be written as follows ($1 \le l, m\le N-1$),
\begin{equation}\label{FG}
 \mathbb{F}_{lm}= \lambda_l \lambda_m \sum_{i=1}^{N}k_i \mathbb{B}_{il}\mathbb{B}_{im} ,\quad\quad 
\mathbb{G}_{lm}= \lambda_l \lambda_m \sum_{i<j=2}^{N}g_{ij} (\mathbb{B}_{il}-\mathbb{B}_{jl})(\mathbb{B}_{im}-\mathbb{B}_{jm}),
\end{equation}
where $\lambda_j\equiv\sqrt{ \frac{\alpha_{1,\ldots,j+1}}{\alpha_{1,\ldots,j}\,\alpha_{j+1} }}$ with \mbox{$\alpha_{1,\ldots,j}=\alpha_1+\ldots+\alpha_{j}$} and where $\mathbb{B}$ is an invertible matrix whose elements can be found on Eq.~(24) of \cite{silv10}.
Let us note that parameters $k_i$ or $g_{ij}$ can be negative, provided all values found for $\omega_i^2$ are positive. 

When $N \le 5$, finding the eigenvalues $d_i$ comes down to solving a polynomial of order ${\cal O}\le 4$, thus analytical expressions for the $\omega_i$ can be obtained. For instance, the complete solution for 3 different particles is given in \cite{silv10}. Analytical expressions for the $\omega_i$ can also be found when all particles are identical \cite{silv10}. 
When the system contains $N_s$ sets of identical particles which interact via two-body forces, another very elegant way to compute the $N$-body problem is presented in \cite{hall79}. In that case, $H_\textrm{ho}$ can be expressed as a sum of Hamiltonians, a term $H_s$ for each set $s$ of identical particles and one term $H_{\textrm{cm}}$ which describes the motion of the centers of mass of the sets of identical particles. All Hamiltonians $H_s$ are completely solvable, and the solutions of $H_{\textrm{cm}}$ are given by Eq.~(\ref{Eohgen}),  meaning that analytical solutions can be found in specific cases such as when $N_s\le5$ or when the total mass of every set is equal. This procedure is generalized in \cite{sema20} for one-body and two-body forces, where an explicit example is calculated for $N_s = 2$. 

In the following, we show that the matrix $\mathbb{J}$ is diagonal, and thus $H_\textrm{ho}$ completely solved, for any number of arbitrary masses provided some specific relations exist between the coupling constants and the masses. 

After some tedious calculations, from Eq.~(\ref{FG}) one can deduce the matrix elements of the symmetrical matrices $\mathbb{F}$ and  $\mathbb{G}$,
\begin{eqnarray}
 \mathbb{F}_{ii}&=&
 \frac{ \left[k_{i+1}\,\alpha_{1,\ldots,i}^2+(k_1+\ldots+k_i)\,\alpha_{i+1}^2\right]}
 {\alpha_{1,\ldots,i}\,\alpha_{1,\ldots,i+1}\,\alpha_{i+1}}\\
 \mathbb{F}_{ij<i}&=&
\Gamma_F(\alpha)\quad
 \frac{(k_1+\ldots+k_j)\,\alpha_{j+1}-k_{j+1}\,\alpha_{1,\ldots,j}}
 {\alpha_{1,\ldots,j+1}\,\alpha_{1,\ldots,i}}\label{Foff}\\
 \mathbb{G}_{ii}&=& 
 \frac{\sum _{m=i+1}^{N} \sum _{l=1}^i g_{l\,m}}{\alpha_{1,\ldots,i}}
 -\frac{\sum _{m=i+2}^{N} \sum _{l=1}^{i+1} g_{l\,m}}{\alpha_{1,\ldots,i+1}}
 +\frac{\sum _{l=1}^{i} g_{l\,i+1}}{\alpha_{i+1}} +\frac{\sum _{l=i+1}^{N} g_{i+1\,l}}{\alpha_{i+1}}\label{Gdiag}
 \\
 \mathbb{G}_{ij< i} &=& \Gamma_G(\alpha) \Bigg[\Bigg.\alpha_{j+1} \alpha_{1,\ldots,i} \left(\sum _{l=1}^j g_{l\,i+1}\right)-\alpha_{i+1} \alpha_{1,\ldots,j} \left(\sum _{l=i+2}^{N} g_{j+1\,l}\right)\nonumber\\
 && +\alpha_{i+1} \alpha_{j+1}\left(\sum_{m=i+1}^{N} \sum _{l=1}^j g_{l\,m}\right) - \alpha_{1,\ldots,i+1} \alpha_{1,\ldots,j} \,g_{j+1\,i+1}\Bigg.\Bigg] \label{Goff}
\end{eqnarray}
where $\Gamma_F(\alpha) = \sqrt{\frac{\alpha_{1,\ldots,j+1}\,\alpha_{1,\ldots,i}\,\alpha_{i+1}} {\alpha_{1,\ldots,i+1}\,\alpha_{1,\ldots,j}\,\alpha_{j+1}}}$ and $\Gamma_G(\alpha) = \frac{
 \sqrt{\frac{1}{\alpha_{1,\ldots,j}}+\frac{1} {\alpha_{j+1}}}
 \sqrt{\frac{1}{\alpha_{1,\ldots,i}}+\frac{1} {\alpha_{i+1}}}
 }
 {\alpha_{1,\ldots,i+1}\alpha_{1,\ldots,j+1}}$. Notice that $\Gamma_F(\alpha)$ and $\Gamma_G(\alpha)$ are strictly positive numbers. We must note at this point that $g_{ij}$ with $i>N$ or $j>N$ are in principle not defined, however in these equations and later in this paper they should be considered as zero.
 
From equations (\ref{Foff}) and (\ref{Goff}) one can notice that the off-diagonal matrix elements of both $\mathbb{F}$ and $\mathbb{G}$ will vanish under certain conditions. In particular, it is easy to see that if $k_i=\rho\,m_i \,\,\forall\ i$ then $\mathbb{F}$ becomes diagonal, and its eigenvalues are all given by $\rho \,m$. 
For $\mathbb{G}$, if $g_{ij}=g_{1j}\frac{\alpha_i}{\alpha_1}$  
then $\mathbb{G}$ becomes diagonal, and its eigenvalues are given by $\frac{\left(g_{1\,i+2}+g_{1\,i+3}+\ldots +g_{1\,N}\right)\,\alpha_{i+1}+g_{1\,i+1}\,\alpha_{1,\ldots,i+1}}{\alpha_1\alpha_{i+1}}$ with $i=1,\ldots,N-1$. The condition $g_{ij}=g_{1j}\frac{\alpha_i}{\alpha_1}$ should not be mistaken for a special requirement on a given particle as the choice of the assignment of particle $1$ is completely free. With this choice of numbering $j$ ($> i$) can take any value from 2 to $N$.

Under these very specific conditions over the nature of the one-body and two-body forces and the masses of the system, we find analytical solutions to  $H_\textrm{ho}$ given by
\begin{eqnarray}
E_\textrm{ho}\Bigr|_{k_i=\rho\,m_i, \,g_{ij}=g_{1j}\frac{\alpha_i}{\alpha_1}}
&=& \sqrt{2} \sum^{N-1}_{i=1} \sqrt{\rho+\frac{\left[\left(g_{1\,i+2}+g_{1\,i+3}+\ldots+ g_{1\,N}\right)\,\alpha_{i+1}+g_{1\,i+1}\,\alpha_{1,\ldots,i+1}\right]}{m\, \alpha_1\,\alpha_{i+1}}}\,\,Q_i.\nonumber\\
\end{eqnarray}

In particular, for the cases where only one-body or two-body forces are present, this gives energies of
\begin{eqnarray}
E_\textrm{ho}^{1B}\Bigr|_{k_i=\rho\,m_i\,\,\,\,\,\,\,} &=& \sqrt{2 \rho}\,\sum^{N-1}_{i=1}Q_i,\\
E_\textrm{ho}^{2B}\Bigr|_{g_{ij}=g_{1j}\frac{\alpha_i}{\alpha_1}} &=& \sum^{N-1}_{i=1} \sqrt{\frac{2\left[\left(g_{1\,i+2}+g_{1\,i+3}+\ldots +g_{1\,N}\right)\alpha_{i+1}+g_{1\,i+1}\alpha_{1,\ldots,i+1}\right]}{m\,\alpha_1\,\alpha_{i+1}}}\,Q_i.
\end{eqnarray}

A simpler expression for $E_\textrm{ho}^{2B}$ can be found under more restrictive conditions: if \mbox{$g_{ij}=\beta\, m_i m_j$} where $\beta$ is a constant, then 
\begin{eqnarray}\label{rest}
E_\textrm{ho}^{2B}\Bigr|_{g_{ij}=\beta\, m_i m_j} &=& \sqrt{2\beta M}\sum^{N-1}_{i=1} \,Q_i.
\end{eqnarray}

Relation (\ref{rest}) has been used to study the possible existence of a quasi Kepler's third law for quantum many-body systems \cite{sema21}. In fact, many approximation methods rely on analytical solutions of simpler Hamiltonians, such as expansions in oscillator basis \cite{Stancu} or in Gaussian states \cite{Varga}.
Furthermore, the existence of analytical solutions for the Hamiltonian $H_\textrm{ho}$ is at the heart of the envelope theory \cite{sema20,hall80} used to solve general translation-invariant $N$-body Hamiltonian \cite{sema20}. 
So, it is particularly relevant to study and expand the availability of analytical solutions of $N$-body Hamiltonians.

\begin{acknowledgments}
This work was supported by the Fonds de la Recherche Scientifique - FNRS under Grant Number 4.45.10.08. 
\end{acknowledgments}

\end{document}